# THE MACROECONOMIC IMPACTS OF ENTITLEMENTS

First Name: Dr. Ateeb Akhter Shah, Last Name: Syed[1,2] | First Name: Dr. Kaneez, Last Name: Fatima[3] | First Name: Riffat. Last Name: Arshad[4]

*Abstract*:

The worries expressed by Alan Greenspan that the long run economic growth of the United States will fade away due to increasing burden of entitlements motivated us to empirically investigate the impact of entitlements of key macroeconomic variables. To examine this contemporary issue, we estimate a vector error-correction model is used to analyze the impact of entitlements on the price level, real output, and the long-term interest rate. The results show that a shock to entitlements leads to decrease in output and lends support to the assertion made by Alan Greenspan. Several robustness checks are conducted and the results of the model qualitatively remains unchanged.

## KEYWORDS

Entitlements, Federal Reserve assets, real output, vector error correction, United States

## JEL CLASSIFICATION

B22, E62, H53, I38

---

[1] Research Department, State Bank of Pakistan, Karachi, Pakistan. ORCID ID: 0000-0003-3872-9609. Email: syed.ateeb@wmich.edu

[2] Department of Economics, Western Michigan University, Kalamazoo, MI, USA

[3] Institute of Management Sciences, University of Balochistan, Sariab road, Quetta, Pakistan. Fatima.sohail@hotmail.com

[4] Department of Economics, University of Balochistan, Sariab road, Quetta, Pakistan. riffatnaseer@hotmail.com

## 1. Introduction

Since the end of the great moderation era, the slowing down of the economic growth in recent years is appearing as a defining challenge for macroeconomists. In the United States (U.S.), this slowdown is generally attributed to increased spending on entitlements. The increasing share of entitlements in GDP of the U.S. has been a matter of serious concern for economists and policy makers. Since last two decades, numerous experts are expressing concerns by highlighting that the Americans are turning into a nation of takers and the Government of the U.S. has become an entitlements Machine Eberstadt (2012). This disproportionate growth in entitlements is viewed as unsustainable and would have serious repercussions for fiscal balances and economic growth.

The former Federal Reserve chairman, Ben Bernanke warned lawmakers that they need to fix the entitlement system. He pointed out that if early and meaningful action is not taken, the U.S. economy could be seriously weakened, with future generations bearing much of the cost. By attributing the rising burden of entitlements on budget to aging population in the United States, he emphasized that if difficult choices are not made, financial stability and healthy economic growth will be lost Bernanke (2010). Similar fears were expressed by Alan Greenspan, the former Federal Reserve Governor. While discussing the causes and implications of secular stagnation at the conservative American Enterprise Institute, he pointed out that spending on entitlements is crowding out gross national saving leading to decreased investment and low productivity Greenspan (2016). He warns recently in an interview that though in the short run economy looks reasonably good, the gains in the economy are draining out from the increasing entitlements and in the long run economic growth will fade away due to crowding out of capital investment Cox (2019). Like Bernanke, he attributes the rise in entitlements to aging population. Samuelson (2016), a renowned columnist highlights the thoughts of Greenspan and emphasizes repeatedly to control over growing burden of entitlements arguing that entitlements are draining funds productivity-enhancing investments.

The term entitlements are referred to benefits that are conferred on any person or unit of government that meets the eligibility requirements established by legislation (Congressional Budget Reform and Impoundment Control Act of 1974) History Art & Archives United States House of Representatives (2020). Social security, Medicare and Medicaid compose more than 75% of entitlement programs in the United States, therefore they are synonymously used as the term entitlements. As pointed out earlier, economists, researchers, analysts and politicians are showing great concern over the rising entitlements spending which is staining the budget, deficit is on the rise since several years. This situation is projected to turn into a nascent fiscal crisis and will fade away economic growth amid pandemic

The share of entitlements spending in GDP is growing year after year and is projected to increase further. As reported in CBO report 2020, mandatory spending on social security and major health care programs make 10.3 percent of the GDP which is projected to be 12.4 percent of GDP by 2030.The budget deficit is projected to climb to 5.0% of GDP in 2030 from 4.6 percent of GDP in 2020. Because of the large deficits, federal debt held by the public is projected to grow, from 81 percent of GDP in 2020 to 98 percent in 2030. The economic growth is projected to slow down. From 2021-2030, output is projected to grow at an average annual rate of 1.7 % CBO report (2020).

Based on the above concerns and statistics, it appears that there is little doubt over the fear that entitlements are one of the major responsible factors for low economic growth and there should be cuts on entitlement expenditures or increase in taxes. However, Lawrence Summers (2013), the former U.S. treasury secretary warned the world about the secular stagflation in a seminal speech at the International Monetary Fund headquarters in New York. He suggested that the solution is increased government spending Summers (2013). Further, Summers (2020) asserts, "We don't need fewer entitlements for the American middle class. We need more". He argues that several other structural factors are responsible for slowing economic growth rather than increased entitlement spending. He explains that secular stagnation is due to persistent low interest rate, global in-equality, and low productivity growth. He believes that real interest rate required to achieve full employment level is so far into negative territory which is effectively impossible. Hence, there is little room for monetary policy to be effective to increase investment and growth. He prescribes that more social insurance will enhance the demand in the economy and increase in interest rate will push the economy forward and contribute to financial stability.

Despite the serious concerns expressed on the role of entitlements in slowing down the growth, the recent literature provides little empirical evidence. Some studies examine the fiscal context of major entitlement programs such as social security, Medicare and Medicaid and recommend reforms in these programs by cutting these spending or increasing tax revenues [see for instance, Palmer (2006) and Gist (2007)]. The increased entitlements are due to aging population in the United States. Maestas, Nicole, Kathleen Mullen, and David Powell (2016) estimate the economic impact of aging population in the states of U.S. over the period 1980 to 2010. They document that 10% increase in the fraction of population ages 60+ decreases per capita GDP growth by 5.5% of which two-third is due to slower growth in labor productivity whereas one-third is caused by slower labor force growth. They predict that annual GDP growth will slow down by 1.2 percentage points this decade and 0.6 percentage points in the next decade due to population aging. However, Acemoglu, Daron, and Pascual Restrepo (2017) investigated the impact of aging on economic growth across countries including OECD economies and conclude that the data do not verify the negative impact of aging population on economic growth. On the other hand, Sheiner (2018) analyzed the long-term impact of aging on the federal budget and showed that debt path of the United States is not sustainable. The literature though limited focuses on the impact of aging population on debt, budget deficit and economic growth. There is no empirical research which investigates the direct impact of entitlements on economy such as economic growth, interest rate and the price level.

The relationship between spending on entitlements and slow economic growth is yet to be empirically investigated as stated by Robert J. Samuelson, "What other economists will think remains to be seen. An obvious question is whether Greenspan's relationships are correlations, not cause and effect" Samuelson (2016). Hence, this article is an attempt to address the question raised by Samuelson. It will contribute to the existing literature in the following way; to best of our knowledge, it will be the first study to use a monthly dataset to empirically investigate the macroeconomic impact of entitlements in the U.S. using a VECM framework.

## 2. Data Description

The data for all the variables used in our paper is taken from Federal Reserve Bank of St. Louis Database (FRED). The main variable of interest is the entitlements. We use the seasonally adjusted personal current transfer receipts, government social benefits to persons measured in billions of U.S. dollars as our measure of entitlements (E). As our sample runs through the time of the great financial crisis, we use the Federal Reserve's assets (A)[5] as a measure of monetary policy. We add A in our analysis to control for the monetary stimulus as it also has the ability to stimulate

[5] For a detailed know-now on the unconventional policy approaches at the zero-lower bound, see Clouse et al. (2003), Bernanke and Reinhart (2004), Bernanke et al. (2004) and Kuttner (2018)

the economy in the short-run. We follow Wheeler (1999) in selection of variables that represent macroeconomy of the U.S., that is, the price level, real output and a long-term interest rate. For this purpose, seasonally adjusted industrial production is included as a measure of real output (Y), seasonally adjusted consumer price index, all items in the U.S. is included as a measure of the price level (P) and the 10-year government bond yields in included as a measure of interest rate (R). All the variables except the interest rate are in log-levels. The interest rate is in levels. The variable A is available only on a weekly basis; therefore, monthly average was taken to form the monthly variable. A is also not seasonally adjusted at the data source; therefore, it is seasonally adjusted using the U.S. Bureau of Census X11 seasonal adjustment program in statistical analysis software (SAS).

## 3. Empirical Analysis

In our paper, we have considered two sample[6] periods for empirical analysis. We start our analysis by testing each variable for stationarity. For this purpose a series of Dickey and Fuller (1981) unit-root tests (ADF) were conducted on each variable. The results of these unit-root tests in presented in Table 1. Introduced by Schwarz (1978), we use the Bayesian information criteria (BIC) to determine the optimal lags to be used in the ADF tests. Panel A of Table 1 shows the results of the ADF test with drift. The test reveals that variables E, P, A and Y are non-stationary in log-levels. R is found to be non-stationary in levels. However, all these variables are stationary in first difference.

Panel B of Table 1 shows the results of the ADF test with trend and drift. Similar to the drift test, the test reveals that variables E, P, A and Y are non-stationary in log-levels. R is found to be non-stationary in levels. However, all these variables are stationary in first difference. All these variables are non-stationary in log-levels/levels; therefore, they should be first differenced to achieve stationarity. However, Engle-Granger (1987) points out that a vector autoregressive (VAR) model estimated in log-levels/levels is miss-specified if the variables are cointegrated. Therefore, we tested the data for presence of cointegration due to Johansen (1988) and Johansen and Juselius (1990). Johansen's [(2000), (2002)] small-sample correction is also employed. The results of the cointegration analysis were obtained using CATs in RATs, version 2 (2005).

BIC is used to determine the optimal lags for cointegration test. BIC suggested to use 6 lags for the test cointegration. The results of cointegration trace test are presented in Table 2. The trace statistic shows that one cointegrating vector exists among the variables. Therefore, to evaluate the impact of E on the P, Y and R, we employ a VECM model estimated using the Engle-Granger (1987) two-step estimator.

We estimate the VECM containing variables E, P, A, Y and R. To preserve degrees of freedom, a maximum lag length considered is 6 lags for the model. The pre-sample extends from 2002M12 to 2003M6, and the estimation of the VECM models is carried out over 2003M7 to 2019M06. The lag length for the model is chosen with the BIC. The BIC suggested a lag length of 5. The residuals from each VECM equation are required to be white noise. To test for serial correlation among the residuals from each VECM equation, we conducted a series of Ljung-Box (1978) tests with the null hypothesis of no autocorrelation. The Q-statistics show that the residuals from each VECM equation in the model estimated with 3 lags are white noise.

---

[6] The first estimation sample is from 2003M7 to 2019M6 (henceforth, short sample). The pre-sample for lags in the ADF tests and VECM for short sample extends from 2012M12 to 2003M6 (1 lag for the difference and 6 maximum lags for the estimation). The second estimation sample is from 2004M2 to 2019M6 (henceforth, long sample). The pre-sample for this exercise extends from 2012M12 to 2004M1 (1 lag for the difference and 12 maximum lags for the estimation). To conserve space, we only report the results of model using the short sample. The results qualitatively remain the same when long sample[6] is used for the analysis.

**TABLE 1     Results of Augmented Dickey Fuller Test (estimation period: 2003M7 – 2019M6)**

| Panel A: Augmented Dickey Fuller Test with drift | | | | | |
|---|---|---|---|---|---|
| Variable | Log-Levels/levels | No of Augmenting Lags (Log-Levels/Levels) | First Difference | No of Augmenting Lags (First Difference) | Critical Value* |
| E | -1.31 | 3 | **-11.43** | 2 | -2.88 |
| P | -1.87 | 2 | **-9.09** | 2 | -2.88 |
| A | -1.24 | 2 | **-9.24** | 1 | -2.88 |
| Y | -2.49 | 4 | **-3.51** | 3 | -2.88 |
| R | -1.59 | 1 | **-9.78** | 1 | -2.88 |
| Panel B: Augmented Dickey Fuller Test with drift and trend | | | | | |
| Variable | Log-Levels/levels | No of Augmenting Lags (Log-Levels/Levels) | First Difference | No of Augmenting Lags (First Difference) | Critical Value* |
| E | -1.93 | 3 | **-11.46** | 2 | -3.43 |
| P | -2.32 | 2 | **-9.24** | 1 | -3.43 |
| A | -1.86 | 2 | **-9.29** | 1 | -3.43 |
| Y | -3.03 | 4 | **-3.52** | 3 | -3.43 |
| R | -3.32 | 1 | **-9.75** | 1 | -3.43 |

Notes: E: Entitlements, P: Price Level, A: Federal Reserve's Assets, Y: Real Output, R: Interest Rate.
* denotes critical value reported in the table at 5% level of significance; The values in bold shows that the variable is stationary in Log-Levels/Levels or first differences

**TABLE 2     The Cointegration Test Results (short sample)**

| Null Hypothesis | Alternative Hypothesis | Eigen Value | Trace Statistic* | Critical Value** | P-values** |
|---|---|---|---|---|---|
| r = 0 | r = 1 | 0.204 | 79.427 | 69.611 | 0.006 |
| r = 1 | r = 2 | 0.108 | 45.507 | 47.707 | 0.081 |
| r = 2 | r = 3 | 0.080 | 24.523 | 29.804 | 0.185 |
| r = 3 | r = 4 | 0.044 | 10.015 | 15.408 | 0.285 |
| r = 4 | r = 5 | 0.027 | 3.713 | 3.841 | 0.054 |

* the small sample corrected trace statistic
** the critical value at 5% level of significance and the p-values are approximated using the gamma distribution, see Doornik (1998)

The results of the model are presented in terms of the impulse response functions (IRFs). To compute IRFs, the residuals from the VECM must be orthogonalized. One technique to compute orthogonalized residuals is Choleski decomposition of contemporaneous relationships. Under the Choleski decomposition, variables in the system are required to be ordered in a particular manner. Variables higher in the ordering contemporaneously influence the variables lower in the ordering and not vice versa. We use the Choleski decomposition with ordering E, A, P, Y and R for our model, when short sample is used.

The current study is concerned with the response of the P, Y and R to a shock in E. This is consistent with the hypothesis we are testing. An increase in entitlement spending may act as a fiscal stimulus and impact the macroeconomy. Hence, it is placed first in the ordering. The key element of our ordering is positioning of policy variables E and A first in the ordering. This permits shocks to these variables impact other variables in the system with in the same month. However, these variables have an impact on E and A through the lags in the system. Hence, like Wheeler (1999), we assume that the policymaker's information set only contains lags of P, Y and R.

In the short-run, certainly contemporaneously, prices are sticky. Hence, Y shocks do not have an impact on P. This places P above Y. Assuming markets are efficient and interest rates reflect all the available information quickly, R is placed at the end. P, Y and R are placed below E and A, the placement of these variables relative to each other is a matter of indifference as long as we are testing the impact of E on Y, P, and R. The conclusion of the model does not change when we alter the ordering of P, Y, and R relative to each other or use alternate ordering.[7] Therefore, we report results with ordering E, A, P, Y, R.

In order to determine if the lag 6 of each variable in the system enters the equations of each variable in the system significantly, we performed a joint significance test on coefficients of this lag. we found that the coefficients on this lag is jointly insignificant. This provides further evidence that lag 5 is appropriate. However, as a robustness check, we also estimated the base model with 6 lags. The results obtained from this robustness check exercise qualitatively remained the same.

## 4. Results and Discussion

The results[8] of our model are presented in Figures 1 through 3. Figure 1 indicates a shock to E has a positive and significant impact on P at first 4 forecast horizons. It also produces a negative and significant impact on P at forecast horizons 6 and 7. Hence, we find that entitlement expenditure is initially inflationary. We know that the funds transferred from the U.S. government to the economic agents leads to an increase in demand; therefore, it will lead to an increase in the price level. However, this impact dampens and eventually becomes insignificant forecast horizon 8 onwards.

---

[7] The alternate-ordering for the model estimated using 3 lags is: E, A, Y, P, R; The difference in this ordering is the placement real output before the price level and this exercise is performed as a robustness check. The main result of the model qualitatively remains the same. In addition to this robustness check exercise, we performed many other robustness exercises and found that the results remained qualitatively the same. These include: estimation of the model over the longer sample period, change in the ordering of the variables in the Choleski decomposition for the long sample, joint significance test on lags 5 through 12 for the long sample and estimation of the model with 6 lags [we do not use more than 6 lags for the long sample; 1) to match with the estimation exercise performed on the short sample, that is, estimation with 5 and 6 lags, 2) due to degrees of freedom that occurs due the use of a longer lag length].

[8] The results of this paper are reported in terms of IRFs. To compute the confidence bands for the IRFs, MONTEVECM procedure is followed in RATs program. The confidence intervals for the IRFs are computed via ten thousand Monte Carlo draws. For each IRF, the bootstrapped confidence bands indicate the 0.05 and 0.95 percentiles of the draws.

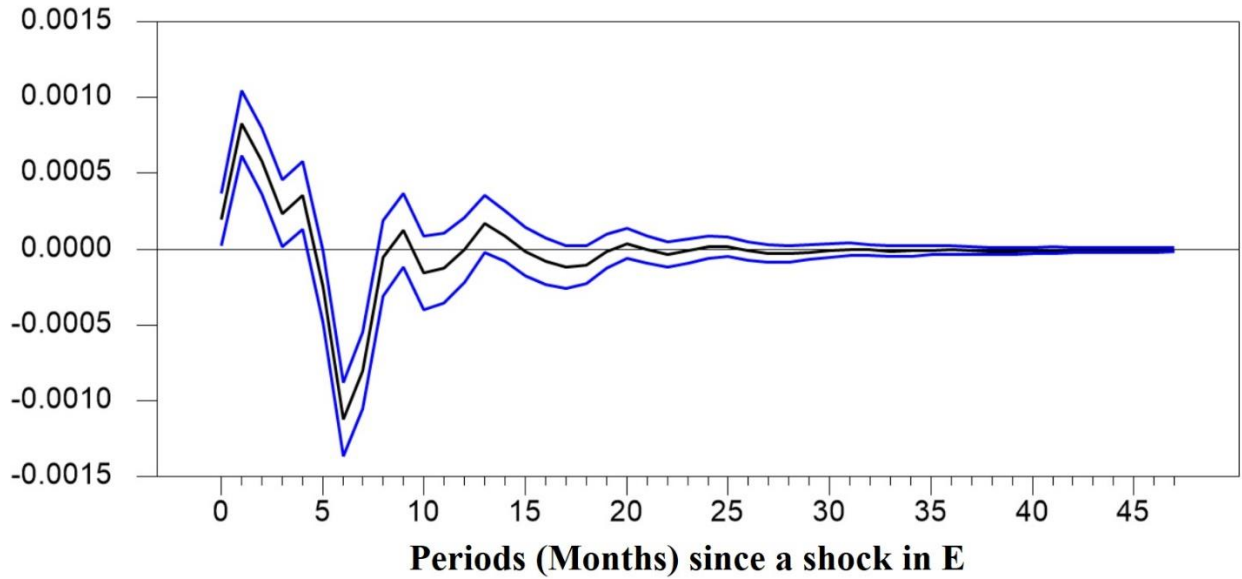

FIGURE 1     Response of the price level to a shock in Entitlements

The main emphasis of our paper is to find out, if the prosperous performance of U.S. fades away due to an increase in E. Hence, Figure 2 contains the most important results of our paper. An analysis of Figure 2 reveals that a shock to E has a short lived positive and significant impact on Y at forecast horizons 1, 2 and 5 months. However, shock to E has a negative and significant impact on Y at 14 forecast horizons (4, 7, 8, 10 to 16, 18 to 20, 22, 24 and 30 months). This result supports the concerns that majority of the experts of the U.S. who believe that economic performance achieved post great financial crisis will eventually start to fade due to increasing burden of E.

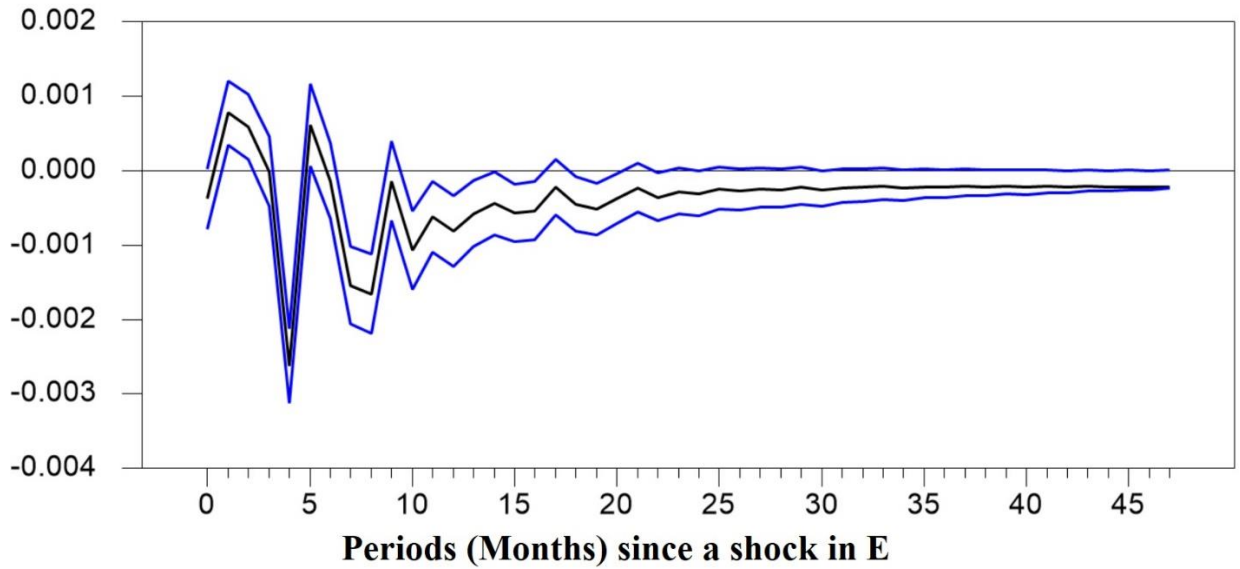

FIGURE 2     Response of real output to a shock in entitlements

Figure 3 indicates that a shock to E has a positive and significant impact on R at first horizons 1 and 13. It also has a negative and significant impact on R at forecast horizon 6 to 8 and is insignificant otherwise. E are financed by tax money as well as through borrowing. Therefore, it may be the case that an increase in E led to an increase in demand for loans through the banking system and pushes R upward. This result is in line with a commonly known notion of the crowding out effect, where a fiscal stimulus if financed by bank borrowing pushes the interest rates high enough that crowds outs the private sector borrowing through this channel. However, the impact becomes negative for a few months and insignificant there onwards.

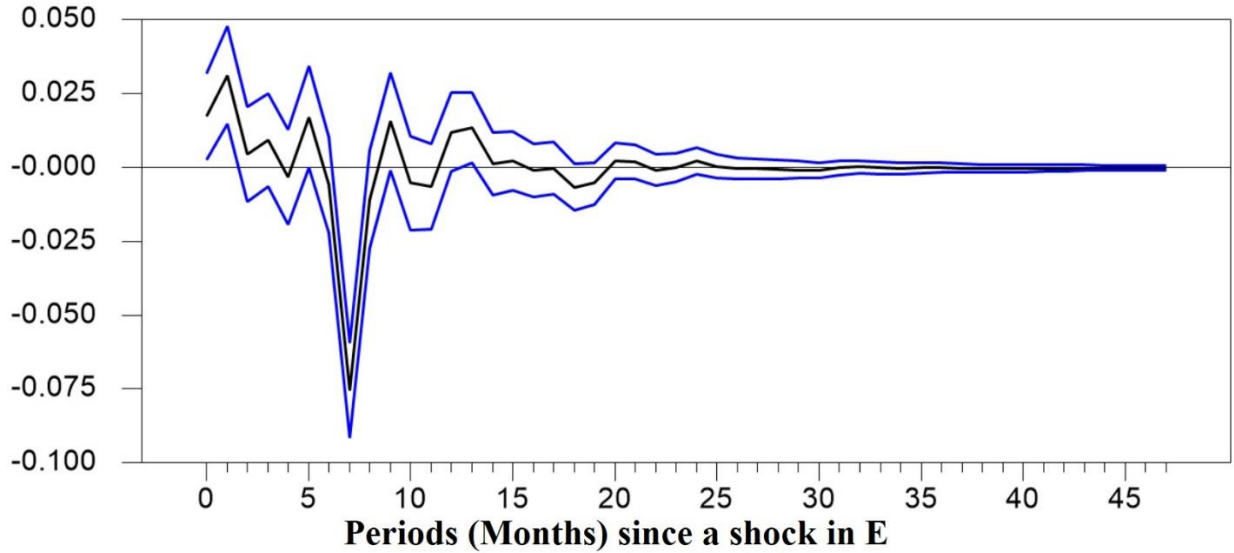

FIGURE 3    Response of 10-year government bond yield (interest rate) to a shock in entitlements

### 5. Conclusion

In this paper, we estimate a VECM to investigate the impact of entitlements on key macroeconomic variables of the U.S. economy. we report results in the form of impulse response functions. The IRFs from the base model reveal that a shock to entitlements has positive and short lived negative, and significant impact on the price level and the interest rate. whereas output is impacted negatively and significantly at 14 forecast horizons. These results remain robust to a change in ordering of the variables in the Choleski decomposition, change in the lag length of the VECM from 5 to 6 lag and a change in the sample (short and long). Hence, this investigation provides empirical evidence on the statement made by Alan Greenspan that though in the short run economy is doing well; however, entitlements will pull healthy performance of economy down. Thus, to move positively on growth trajectory, the U.S. must manage entitlements smartly.